\documentclass[12pt,preprint]{aastex}

\slugcomment{Accepted to AJ}
\shortauthors{Howell et al.}
\shorttitle{IR Spectroscopy of VV Pup}

\begin{document}

%
%

\title{Low State, Phase-Resolved IR Spectroscopy of VV Puppis}

\author{Steve B. Howell\altaffilmark{1}, 
Thomas E. Harrison\altaffilmark{2},
Ryan K. Campbell\altaffilmark{2},
France A. Cordova\altaffilmark{3},
\and
Paula Szkody\altaffilmark{4} 
}
\altaffiltext{1}{WIYN Observatory and National Optical Astronomy Observatory,
950 N. Cherry Ave, Tucson, AZ 85719 {\it howell@noao.edu}}
\altaffiltext{2}{Department of Astronomy, New Mexico State University,
Box 30001, MSC 4500, Las Cruces, NM 88003 
{\it tharriso@nmsu.edu}}
\altaffiltext{3}{Institute of Geophysics and Planetary Physics, Department of
Physics, University of California, Riverside, CA 92521
{\it france.cordova@ucr.edu}}
\altaffiltext{4}{Department of Astronomy, University of Washington,
Box 351580, Seattle, WA 98195
{\it szkody@astro.washington.edu}}
\keywords{Infrared Spectroscopy -- Stars: individual (EF Eri, VV Pup, EQ Cet)}

\begin{abstract}
We present phase-resolved low resolution $JHK$ and higher resolution $K$-band spectroscopy of the
polar VV Pup. All observations were obtained when VV Pup was in a low accretion state
having a K magnitude near 15. 
The low resolution observations reveal cyclotron emission in the $J$ band
during some phases, consistent with an origin near the active 30.5 MG pole on the white 
dwarf. 
The secondary in VV Pup appears to be a normal M7V star and we find that the
$H$ and $K$ band fluxes are entirely due to this star at all
orbital phases during the low accretion state.
We use our higher resolution Keck spectroscopy to produce the first $K$-band radial velocity
curve for VV Pup. Our orbital solution yields $K_2$=414$\pm27$ km sec$^{-1}$ and leads to mass
estimates of M$_1$=0.73$\pm$0.05 M$_{\odot}$ and M$_2$=0.10$\pm$0.02 M$_{\odot}$.
We find that the mass accretion rates during the normal low states of 
the polars VV Pup, EF Eri, and EQ Cet are near 10$^{-13}$ M$_{\odot}$ yr$^{-1}$.
The fact that \.M is not zero in low state polars 
indicates active secondary stars in these binary systems, 
including the sub-stellar donor star present in EF Eri.
\end{abstract}

\section{Introduction}

VV Puppis is considered to be a typical magnetic cataclysmic
variable or ``polar''. It contains a highly magnetic white dwarf and a late-type
secondary star. VV Pup has an orbital period of 100 minutes,
and magnetic field strengths of 30.5 MG and 56 MG at the two white dwarf
poles. As with all polars, VV Pup shows large brightness
variations of a few magnitudes between what are termed
high and low states. These are times of normal and very low mass transfer 
respectively for material accreted from the secondary star to the primary white dwarf.
VV Pup has been studied in detail in the optical 
since the mid-1960's (Walker 1965; Liebert et al., 1978; Schneider \& Young,
1980; Cowley et al., 1982; Wickramasinghe et al., 1989) 
and at high energy
(X-ray, FUV, and UV: see Patterson et al., 1984; Vennes et al., 1995, Imamura et al.,
2000; and Hoard et al., 2002).
Nearly all studies have occurred
during a high state when emission from the accretion stream, the accretion
column near the white
dwarf magnetic pole, and the accretion region at the pole itself 
dominate its flux output.
One recent X-ray study of VV Pup (Pandel and Cordova 2005) was performed during a
low state. At this time, it was observed that the mass transfer occurred in an
irregular fashion and the X-ray flux from the accretion pole region 
varied by more than an order of magnitude during orbital phases when the active pole
was in view. 
These authors concluded that irregular accretion is common in
low-state polars, and that such flares are due to ``accretion rate fluctuations"
representing strongly varying mass transfer from the companion star. 
The flares are likely to result either from
capture of coronal mass ejections from the secondary star by the white
dwarf's strong magnetic field, or from solar flares near the L1 point. This
indicates stellar activity in the secondary star in polars.

VV Pup has been in an extended low state much of the past three years. 
%
%
Based on our stochastic
monitoring efforts, the current low state appears to have started near 
December 2002 and continued througout most of 2005 (until October or November)
with a brief (2-3 weeks?) volley back
to an apparent normal high state in December 2003.
During these past few years, we
have undertaken extensive photometric and spectroscopic studies of VV Pup 
in the optical and IR in order to 
confidently identify the secondary star and assess its properties and to
study the non-accreting white dwarf surface Zeeman field and
low state cyclotron emission. Results of our optical low state
spectroscopy and photometry as well as our magnetic modeling effort
are presented
elsewhere (Mason, et al., 2005, 2006). 
Here we concentrate on IR spectroscopy
and the secondary star.
We present the first low state, phase-resolved $K$-band spectroscopy of VV Pup
wherein we determine an orbital solution for the binary. 
We find the true K$_2$,
the orbital phase of the secondary star at inferior conjunction, and the
spectral type of the secondary star. We are also able to provide robust mass
estimates for M$_1$ and M$_2$.

\section{Observations}
VV Pup was observed during a low state with the NIRSPEC
instrument\footnote{http://www2.keck.hawaii.edu/inst/nirspec/nirspec.html}
on Keck II under photometric conditions on 17 Feb. 2005. 
Table 1 presents a journal of the observations. NIRSPEC was used in low
resolution mode with the slit width set to 0.38". Our grating tilt was chosen
to cover the $K$-band spectral range of 2.04 to 2.46 microns with a dispersion of
4.27 \AA/pixel. This setting provided access to 
secondary star absorption features such as Ca, Na, and CO. We used four minute
exposure times throughout and nodded the star along the slit using the standard
NIRSPEC 2-nod script, obtaining 22 individual exposures. Observations of bright A0V stars taken nearby in time and
location, were used for telluric correction. We reduced the Keck spectra using
the IDL routine
``REDSPEC"\footnote{http://www2.keck.hawaii.edu/inst/nirspec/redspec/index.html}
especially developed for NIRSPEC reductions. REDSPEC averages the two spectra
obtained in the ``A'' and ``B'' slit positions, resulting in eleven processed 
spectra with relative fluxes. A0V stars are nearly featureless
in the $K$-band except for a prominent HI Brackett line at 2.16 microns. The
REDSPEC software does not attempt to correct for this feature but can
interpolate across it thus reducing the impact of the line on the program
object spectrum. Because a weak telluric feature sits near the Brackett line, 
the region very near the 2.16 micron Brackett feature is slightly compromised. 

Figure 1 shows our final summed $K$-band Keck spectrum of VV Pup. We note it shows
what {\it appears} to be a very weak HI Brackett emission feature. We believe
that this emission is probably artificial and is caused by the patching-over of 
the standard star absorption feature during data reduction as noted above.  
We can not, however, completely rule out the 
possibility of residual accretion causing some weak HI Brackett 
emission during this extended low state.
While we obtained phase-resolved spectroscopy (see Table 1) we
noted no significant spectral differences during the single orbit 
that we observed other
than radial velocity changes in the absorption lines.
Figure 1 was produced by correcting each of our 11 spectra for the
secondary star motion using the radial velocity result below. 
The final summed spectrum reveals that the
secondary star in VV Pup appears to be a normal M7V star (also see Harrison et
al., 2005).

We also observed VV Pup using SPEX on the IRTF in ``Prism mode'', from 9:58 to
11:42 UT on 2005 February 7. The dispersion in this mode is 34 \AA/pix, and
the spectra cover the wavelength range 0.64 to 2.55 $\mu$m. The integration 
times for all exposures were four minutes, and we obtained 20 spectra. The spectra were reduced using
SPEXTOOL (Vacca et al. 2003). A nearby A star was used to remove telluric 
features from these spectra. Figure 2 presents our IRTF/SPEX phase-resolved 
data for VV Pup covering the 
$IJHK$ bands. Each spectrum plotted in this figure is the average of two 
individual spectra (see Table 2). The S/N of these data are not high, 
as VV Pup is a challenging
target for a 3-m telescope, even at the lowest resolution setting of SPEX and
the ``smoothed" appearance of the data is solely due to its limited 
spectral resolution. It 
is clear, however, that at least one cyclotron
hump is present (near 1.2 $\mu$ m), which appears to be the $n$ = 3 harmonic
from the 30.5 MG pole. The reality of the
features blueward of $\sim$0.9 $\mu$m are somewhat difficult to ascertain. There are 
several telluric features in this region, and due to the target faintness, 
they are difficult to properly remove. In addition, the Paschen 
continuum occurs in this wavelength region
(terminating at 0.82 $\mu$m), as well as the likely contribution of the $n$ = 4 
harmonic from the 30.5 MG pole (centered near 0.90 $\mu$m). The $H$ and $K$
bands appear to be completely dominated by photospheric flux 
from the secondary star during
all orbital phases. Near phase 0.5, the $J$-band spectra is free of cyclotron
contributions and appears to be
consistent with a normal cool late M-type secondary star. 

\section{Radial Velocity Fitting}

Using our Keck phase-resolved spectroscopy we measured the line centers of the Na
doublet (2.2062 \& 2.2090 microns), the Ca triplet (2.2614, 2.2631 \& 2.2657
microns) and the blue edge of the first CO band-head (2.2935 microns).
We fit each of the lines in all eleven spectra (see Table 1) 
after first boxcar smoothing them by 5 
to allow robust line fitting. Figure 3 shows an example of one of our Keck spectra,
the region around the Na line, and a smoothed version of this same region. 
IRAF was used to
fit each spectral line of interest (the CO edge was fit by eye) 
with a single Gaussian profile returning a line
center and fitting uncertainty. In most cases, the
Gaussian line fitting determined the line center to one-fifth of the (binned) 
spectral resolution, or $\pm$5\AA. 
The line center uncertainties were noted for each
measurement and these errors properly propagated through the
analysis discussed below.

The CO band-head edge fitting was attempted based on some success producing 
a radial velocity
curve with this technique as seen in Howell et al. (2000). For VV Pup
these measurements proved to be consistent with, but much less precise than, those
described below for the Na I lines. Use of the CO measurements simply added noise to
the final fit of a RV solution so in the end we dropped their use. 
Likewise, the Ca triplet line centers were measured but proved to be chaotic
in their velocity determinations. The cause is due to the redward 
component of the unresolved triplet changing strength relative to the 
other two components due to its location on the rapidly 
falling continuum (caused by H$_2$O steam band absorption in the secondary) 
and we again dropped these measurements from the final fitting.

Measurement of the strong Na I doublet produced highly reliable velocity
determinations with an essentially constant velocity uncertainty 
of $\sim$0.005 microns (the resolution element), or
$\sim$50 km sec$^{-1}$. The use of this doublet was also seen to produce good
results for the polar ST LMi in Howell et al. (2000). 
The Na I doublet is not resolved in our observations and we 
fit the entire absorption line taking the Na I doublet line strength
weighted center to be 2.2076 microns. This line center value will 
have only a negligible effect
on the radial velocity solution values for K$_2$ or the red-to-blue crossing 
phase but could effect our measured systemic velocity determination.

We determined the orbital phase of each of the eleven
observations listed in Table 1 using the photometric ephemeris presented in 
Walker (1965). Walker's ephemeris for the time of high state
maximum light is 
$$ JD~~ 2427889.6474 +0^{d}.0697468256 E $$ and was based on a number of nights of
multi-color 1P21 PMT photometric data obtained at the Lick 120-inch telescope.
We note here that Walker (1965) lists his ephemeris as JD 
but calculations using his published tables of data, confirm that it is 
really HJD\footnote{A fact also noted by Schneider \& Young (1980).}. 
Additionally, the listed ephemeris in Diaz \& Steiner (1994, their eq.
1) is incorrect while Patterson (1984) suggests a correction 
of $\sim$0.03 cycles to
Walker's photometric value due to timing measurements that suggest 
a small drift has occurred.
While Walker (1965) is a fairly old maximum light photometric ephemeris, 
it has not only been used by all VV Pup
investigators to date and, as we will see, it remains highly accurate. The time 
of high state optical maximum light (Walker's phase 0.0) should be the time 
when the observer is looking
most directly at the active accretion region on the white dwarf. 
In most polars, the active pole is thought to be locked 
near an {\it orbital phase} of $\phi$ = 0.75 $- $ 0.85, 
where phase 0.0 is defined as the secondary star inferior conjunction.

Taking our velocity measurements for the Na line listed in Table 1 and using the
orbital phases calculated from the Walker photometric ephemeris, we produce the radial
velocity curve shown in Figure 4. Our best fit sine-curve solution is also plotted
and the parameters of this fit are listed in Table 3. We note that this is the
first unambiguous spectroscopic orbital solution for VV Pup. 
The radial
velocity curve shows a {\it blue-to-red} crossing at orbital phase 0.48 
placing the
{\it red-to-blue} crossing of the secondary star inferior conjunction at 
orbital phase $-$0.02 or
very near Walker's photometric phase 0.0. Hoard et al. (2002) 
also use Walker's ephemeris and,
based on high state FUV observations, conclude that the inferior 
conjunction should
occur near phase 0.1. Allen and Cherepashchuck (1982) suggest that true orbital
phase zero should occur at Walker's phase 0.05$\pm$0.03 based on $J$-band photometric 
observations of 
secondary star ellipsoidal variations made during a low state (J$\sim$15). 
However, we note above (and will detail in the Discussion) 
the fact that during a low state, the
$J$-band flux for VV Pup is highly modulated by cyclotron emission. The 
assignment of low state polar
light curve modulations caused by cyclotron emission 
to ellipsoidal variations has occurred before (see Howell et al., 2001).
We believe that the Allen and Cherepashchuck
results are such a case and the modulation they observed was entirely due to 
orbitally modulated $J$-band cyclotron emission and not ellipsoidal variations.
Given the previous work using high state
optical emission line decomposition and high energy observations, as well as 
the time since Walker's
determination, we take these literature values to agree 
within the uncertainty of the
original ephemeris. Our uncontaminated view of the secondary star provides the
best yet determination of the true orbital ephemeris and we use our
best fit to to 
determine a spectroscopic {\it red-to-blue} crossing ephemeris for VV Pup
of 
$$ HJD~~ 2453418.93818(\pm0.00005) +0^{d}.0697468256 E $$ 
where orbital phase 0.0 is the secondary star inferior conjunction.
Our solution has an uncertainty of about 
$\pm$0.01 in phase for the time of zero crossing.

\section{Discussion}

\subsection{Low-Resolution Spectroscopy and other Low State Polars}

Our phase-resolved, low-resolution spectroscopy of VV Pup is presented in
Fig. 2. The data are fairly noisy due to the extreme faintness of VV Pup. For
most of the orbit, the $JHK$ data are consistent with the spectrum of a 
late-type star. However, during the orbital phase 
interval $\sim$0.8 to 0.1, two broad humps, 
centered near 0.9 and 1.2 $\mu$m, develop in the $J$ band. These humps are
strongest and most easily seen in our phase 0.00 and 0.09 spectra.
and are caused by cyclotron radiation from one of
the magnetic poles of the white dwarf in VV Pup. 
As described above, the feature at 1.2 $\mu$m is consistent
with the $n$ = 3 harmonic from the 30.5 MG pole. 
The cyclotron ``continuum" consists of an optically thick Rayleigh-Jeans 
tail at the low harmonics (long wavelengths) and a cyclotron hump modulated power law
for higher harmonics (shorter wavelengths). 
As the harmonic number increases, the harmonic structure becomes
optically thin causing the discernible ``cyclotron humps" 
that modulate the continuum flux. 
The $n$ = 2 harmonic from this same pole would be located at
1.78 $\mu$m, but this region is compromised by telluric water vapor absorption 
and early
cyclotron harmonics are generally optically thick providing only 
a possible weak 
continuum contribution (Wickramasinghe \& Ferrario 2000). 
We see no evidence for the $n$=2 harmonic in our data.

The development of 
the 0.90 $\mu$m ($n$=4) feature is quite rapid, and
appears in the spectrum at the same time as the 1.2 $\mu$m feature. 
Mason et al., (2005, 2006) present high S/N optical spectroscopy of VV 
Pup during this same low
state, and their results show unambiguous evidence for the 30.5 MG pole to 
produce strong
cyclotron harmonics ($n$=5,6) in the red (having widths near 800-1000 \AA)
which rapidly appear during approximate orbital phases 0.8 through 0.1
and disappear during the remainder of the orbit. Their results are in 
general accord 
with ours. Furthermore, Mason et al. note no cyclotron emission
from the 56 MG pole, which we do not see as well, revealing that 
any residual mass accretion
during the low state only occurs at the lower strength (30.5 MG) 
white dwarf magnetic pole that is
energetically closer to the secondary star.


The cyclotron harmonics observed in VV Pup are weak compared with
other low state polars we have observed using SPEX (Campbell et al. 2005;
also see Harrison et al., 2004). 
Figure 5 shows a single IRTF/SPEX spectrum of the star EQ Cet. 
EQ Cet was studied in the optical by Schwope et al. (1999)
and assigned a magnetic field strength (for the active pole)  
of B$\sim$45 MG based on the single large cyclotron hump ($n$=4) 
observed in the near-IR.
Due to this single dominant cyclotron hump seen in the optical, EQ Cet 
was called a LARP (low accretion rate polar)\footnote{A more recent definition
of the LARPs is provided by Schmidt et al. (2005).}. 
Our SPEX observation shows the presence of three 
cyclotron humps in the $J$-band region 
that can be assigned to the $n$=2,3,4 (blue hump edges 
at 1.4, 0.96, and 0.73 $\mu$m) harmonics 
of a B=37 MG magnetic field. Thus, the once ``exotic" one hump 
LARP, EQ Cet, appears to be a normal looking low state polar when observed 
in the near-IR; the spectral
region for which EQ Cet's magnetic field strength produces more 
than one cyclotron feature.
A detailed discussion of our phase-resolved EQ Cet spectroscopy will be presented in 
Campbell et al. (2005).

EF Eri, a polar with a magnetic field strength of B=16-21 MG, also shows stronger low state 
cyclotron humps in the $J$-band than does VV Pup.
We note that EF Eri has a weaker magnetic field strength than VV Pup 
while EQ Cet is comparable. Cyclotron hump emission strength
mainly depends on the electron temperature (i.e., $\propto$ M$_{WD}^{\sim0.66}$), 
the magnetic field strength ({\bf B}), and the 
(instantaneous) mass accretion rate with lower dependence on which cyclotron 
harmonic and viewing angle
(Lamb and Masters 1979; Wickramasinghe \& Ferrario 2000 ). 
VV Pup is believed to have the most massive white dwarf of the
three systems discussed here (0.65 M$_{\odot}$ compared with 0.6 M$_{\odot}$ for 
EF Eri and EQ Cet), while its magnetic field strength lies between the other two stars. 
How can we account for the relative low state cyclotron emission strengths observed in these
three stars?  Considering the differences in white dwarf mass and {\bf B}
between these three stars and the fact that Pandel and
Cordova (2005) determined that VV Pup has a low state mass accretion rate of 9.4 
$\times$ 10$^{-13}$ M$_{\odot}$ yr$^{-1}$, we can simplistically argue that EQ Cet must
have a low state mass accretion rate of less than or equal to that of VV Pup while
EF Eri's low state \.M must be nearly twice as high.
Schwope et al. (1999) calculated 
that EQ Cet has a low state \.M of 1.5 $\times$ 10$^{-13}$ M$_{\odot}$
yr$^{-1}$, in
agreement with our simple model. EF Eri, on the other hand, seems to have low states
where H$\alpha$ is present and lower states where no emission is observed and the
Zeeman split hydrogen absorption lines are observable (Beuermann, et al., 2000; Howell 2004).  
Additionally, EF Eri's X-ray luminosity (in the lower state) is 7 $\times$ 10$^{28}$
ergs sec$^{-1}$, far below the value of 1 $\times$ 10$^{30-31}$ ergs sec$^{-1}$ 
observed for VV Pup's low state (Pandel and Cordova 2005) suggesting a very low mass
accretion rate. Thus, the level of cyclotron hump strength during a low state may not
be a simple proxy allowing an estimate of the mass accretion rate.

These low state mass accretion rates are 2--3 orders of magnitude below 
that expected for
short-period (below the period gap) cataclysmic variables, polars in a high 
state, and disk accreting dwarf novae. We note that during what appear to be 
normal polar low states, 
the \.M values agree with the value proposed 
as the mass accretion rate in a LARP. 
We note here for interest that the mass loss rate of the sun via the 
solar wind, is 
$3\times$10$^{-14}$ M$_{\odot}$ yr$^{-1}$ and that of a chromospherically 
active late-type main sequence star
can be $\sim$10$^{-13}$ M$_{\odot}$ yr$^{-1}$.
LARPs were originally taken to be polars with very low \.M values during their
low states. A newer definition places them as detached binaries with very cool
white dwarfs believed to be pre-polars (Schmidt et al., 2005).
Cyclotron emission in LARPs is assigned to accretion by the white dwarf of stellar wind
material or flaring events on the secondary. If the same process provides low state
cyclotron emission in normal polars\footnote{Roche lobe overflow is believed to stop
during a low state.}, then the secondary stars must be
chromospherically active. This includes the cool 
(T$_{eff}\sim$1400K), sub-stellar (0.045 M$_{\odot}$) donor object in EF Eri.

\subsection{K-Band Spectroscopy}

Our K$_2$ value, 414 km sec$^{-1}$, is the first value directly measured for
the secondary star in VV Pup. Previous attempts (see Table 4) 
were performed during high states and used the fact that the narrow
or sharp component of optical H and He emission lines form on or near the
secondary star. 
Schneider \& Young (1980) separately measured the broad and narrow 
components of hydrogen
emission lines and used the narrow component velocities to estimate K$_2$.
Cowley et al., (1982) applied a similar technique, producing a rather complex
narrow line radial velocity curve. 
Using new optical spectroscopy, Diaz and Steiner (1994) applied Doppler mapping
techniques to estimate K$_2$. These authors realized that due to the location
of the formation site of the narrow hydrogen emission line components, near L1, 
an inverse K correction (Wade \& Horne 1988) would be needed to convert 
the observed K$_2$ value to 
the true K$_2$ value for the secondary star center of mass.
Diaz and Steiner applied this correction (assuming values 
for $i$, M$_1$, and M$_2$,where M$_2$ was taken to be a
M4-5V star of mass 0.2M$_{\odot}$) as well as averaged their value with previous
estimates to produce their final best estimate for K$_2$.
Table 4 summarizes the K$_2$ history for VV Pup.

All of these studies were undertaken in the optical and during a high state and
thus none of them detected the secondary star directly. The narrow component of
the optical emission lines are formed near L1 and are highly contaminated by
the broader, more variable emission from the accretion stream and column.
Additionally, the secondary star spectral type was assumed to be M4 to M5 based
on seeing apparent TiO bands in the less than ideal low state, low resolution optical spectrum of VV 
Pup presented in Liebert et al. (1978). This secondary star type became
widely accepted as it generally agreed with the JHK color of VV Pup 
and the spectral type expected based on the (fallacious) idea that 
cataclysmic variables have a
``Spectral Type (mass) - Roche lobe radius" relation. Our data eliminate 
this shortcoming via our direct determination of the secondary star spectral
type 
(see Harrison et al., 2005) and our direct measurement of absorption features from
the secondary that provide the proper K$_2$.

Taking 
M$_2$ to be a normal M7V star as discussed above, we can estimate its mass
to be 0.08 M$_{\odot}$ (Golimowski et al.,
1995; Cox 2000). A spectral type of M8V is ruled out, as shown in
Harrison et al. (2005), by a comparison of the Na to CO absorption depths: 
Na decreases in depth while CO becomes
far deeper in later type stars. It is harder to completely rule out an M6V type
(M = 0.12 M$_{\odot}$) for the secondary star in VV Pup, 
while  a M5V has far to weak CO absorption.
The dynamic mass, spectral type relationship found by Kirkpatrick \&
McCarthy (1994) places M6.5V as the spectrally defined boundary between
normal M dwarfs and brown dwarfs. Spectral types later than this boundary value
are confused and may be normal main sequence stars or brown dwarfs (the latter 
being objects
that are degenerate in temperature, mass, and radius). However, only the youngest
brown dwarfs appear similar to late M stars and it is unlikely the secondary star
in VV Pup is young. 

A number of researchers have estimated the binary inclination for VV Pup.
Schneider \& Young (1980) use $i$=66$^{\circ}$ as a best fit value in order to keep
the white dwarf mass within reason. However, they assumed a 0.2 M$_{\odot}$ secondary,
and if their mass function is corrected for our new mass estimate, $i$ becomes
76$^{\circ}$. Cowley et al. (1982) only set a lower limit on K$_2$ and in order to
keep their white dwarf mass near 1.0 M$_{\odot}$, they were forced to set $i$ to 
55$^{\circ}$. Their solution produced a secondary star with a mass of $\ge$0.25
M$_{\odot}$, far to massive to be consistent with our K-band determination. If we
correct their analysis with our low mass secondary, $i$ becomes 75 degrees.
Using polarimetry, Cropper (1988) measured the system geometry for the white dwarf,
solving simultaneously for the location of the magnetic pole and the inclination of 
the binary system. He found a value for $i$ of 76$\pm$6$^{\circ}$.
Sirk \& Howell (1998) used EUV observations of VV Pup during a high state 
and fit their observations with a 3-D accretion region model. Their best fit
binary inclination was 73$\pm$2$^{\circ}$. Finally, Vennes et al. (1995)
modeled the white dwarf flux distribution in an EUV spectrum and found a best fit by
using a corrected Cowley et al. inclination value of $i$=76$^{\circ}$.
Taking the Cropper and Sirk \& Howell results (and the consistent value found by
correcting the older results for the secondary mass plus the value used by 
Vennes et al.), we take the inclination of VV Pup's orbit as 75$\pm$6$^{\circ}$.

Taking M$_2$ to be 0.10$\pm$0.02 M$_{\odot}$ and M$_1$=0.65$\pm$0.5
M$_{\odot}$ (Vennes et al., 1995, Pandel \& Cordova 2005), 
we find $q$ = $\frac{M_2}{M_1}$ = 0.15 (not far from 
0.18 found by Szkody et al., 1983) and would therefore predict a K$_1$
of $\sim$64 km sec$^{-1}$. Taking our $q$ value to be correct and solving the mass
function for M$_1$ (a function with only a weak dependence on $q$), we find
$$ M_1 = (1+q)^2~~  sin^{-3}i~~ P_{orb}~~ {K_2}^3~~ {2 \pi G}^{-1} $$
yields M$_1$=0.73$\pm$0.05M$_{\odot}$ where essentially all of the 
uncertainty is due to our poor knowledge of the binary inclination. 
The value is 
in agreement with that used in Vennes et al., and in Pandel \& Cordova as well
as the mass functions presented in
both of the detailed optical spectroscopic studies of Schneider \& Young (1980)
and Cowley et al., (1982), if M$_2$ is taken to be $\sim$0.10 M$_{\odot}$.

Examination of Fig. 1 shows that during this low state, all of the $K$-band flux is
due to the secondary star. Bailey (1981) used his $K$-band flux 
and, well, ``Bailey's method" 
to determine VV Pup's distance while Vennes et al., (1995) 
modeled EUV observations
of the accretion region on the white dwarf. Both of these authors determined
that VV Pup is about 145 pc away. For a normal M7V star, with 
M$_K$ $\sim$ 10.0, the observed magnitude for this distance should be $K$ =
15.9 for the secondary star alone. It is difficult to extract precise
photometry from either our Keck or IRTF data due to the small slits used, but 
both give $K ~ \approx$ 15.5. Szkody \& Capps (1980) obtained a photometric value of 
$K$= 15.1$\pm$0.2 during the
1978 May low state. Using these K magnitude values and 
if the secondary star is ``normal'', the implied distance is $\approx$ 120 pc.

\section{Conclusions}

We present the first K-band, low state radial velocity solution for 
VV Pup. These observations allow, for the first time, a true orbital phasing for the
binary to be determined. 
We note that during the low state, all the $H$ and $K$ band 
flux is from the secondary star and 
we can use the secondary star's spectral appearance and sodium absorption features
in its photosphere to derive mass estimates for both components of the
binary. 
The secondary star in this system seems to have a mass
(M$_2$=0.10$\pm$0.02 M$_{\odot}$) and luminosity 
which are consistent with a normal, late-type
M7V star. We determine a white dwarf mass of M$_1$=0.73$\pm$0.05 M$_{\odot}$.
Normal secondary stars seem to be the rule for polars but the
exception for the 
remainder of the cataclysmic variables (Harrison et al., 2005).
We also provide an improved orbital ephemeris. Table 5 summarizes our determined
parameters for VV Pup.

During this low state, the cyclotron harmonics in VV Pup were very weak, much
weaker than in the LARP EQ Cet (Campbell \& Harrison 2004), or EF 
Eri during its extended (8+ year) low state (Harrison et al., 2004). 
It seems that 
during low states, the accretion rates in polars can be just
as low as those observed for LARPs ($\sim$10$^{-13}$ M$_{\odot}$ yr$^{-1}$),
values in agreement with single, highly active M stars..
Our findings suggest that low state polar
white dwarfs can remain at a low activity level by accreting material from 
the secondary star stellar wind and/or flare or star-spot activity.
The tidally locked, spun-up secondaries in CVs should be enhanced in activity and capable of easily
providing the needed mass loss levels.
For VV Pup, low state observations indicate that a 
stellar wind and/or chromospheric activity is
still on-going in a M7V star. For the secondary object in EF Eri, 
similar processes need to 
be present even though it is a cool, sub-stellar brown dwarf-like star.

\acknowledgments
We wish to thank John Thorstensen for providing us with his preliminary
distance determination of VV Pup. The anonymous referee is also thanked 
for suggestions and comments that led to a much improved paper.
Data presented herein were obtained at the W.M. Keck Observatory, which is
operated
as a scientific partnership among the California Institute of Technology, the
University of
California and the National Aeronautics and Space Administration. The Observatory
was made
possible by the generous financial support of the W.M. Keck Foundation.
The authors wish to recognize and acknowledge the very significant cultural role
and
reverence that the summit of Mauna Kea has always had within the indigenous
Hawaiian
community.  We are most fortunate to have the opportunity to conduct
observations from this mountain.

{\bf Note added in proof:} 
Thorstensen (2005, private communication) has measured a preliminary 
parallax $\pi_abs = 8.6 \pm 1.5$ mas for VV Pup. A Bayesian distance
estimate taking the proper motion and Lutz-Kelker bias into account
gives 129(+29,-21) pc, in close agreement with the distance found 
here.

\newpage

\begin{deluxetable}{cccc}
\tablenum{1}
\tablewidth{5.4in}
\tablecaption{Keck Spectra of VV Pup}
\tablehead{
 \colhead{Spectrum}
 & \colhead{HJD of mid-exposure}
 & \colhead{Orbital phase\tablenotemark{a}}
 & \colhead{Velocity (km sec$^{-1}$)}
}
\startdata
\hline
1 & 2453418.85622 & 0.8205812 & -460.7 \\
2 & 2453418.86238 & 0.9089006 & -298.9 \\
3 & 2453418.86958 & 0.0121311 & 8.2 \\
4 & 2453418.87587 & 0.1023145 & 146.8 \\
5 & 2453418.88265 & 0.1995232 & 266.4 \\
6 & 2453418.88875 & 0.2869824 & 294.9 \\
7 & 2453418.89555 & 0.3844778 & 28.5 \\
8 & 2453418.90158 & 0.4709334 & -82.9 \\
9 & 2453418.90816 & 0.5652746 & -322.9 \\
10 & 2453418.91423 & 0.6523036 & -478.3 \\
11 & 2453418.92842 & 0.8557538 & -496.0 \\
\hline
\enddata
\tablenotetext{a}{Determined using the photometric ephemeris given 
in Walker (1965).}
\end{deluxetable}{}

\begin{deluxetable}{ccc}
\tablenum{2}
\tablewidth{3.5in}
\tablecaption{IRTF Spectra of VV Pup}
\tablehead{
 \colhead{Spectrum}
 & \colhead{UT of mid-exposure}
 & \colhead{Orbital phase\tablenotemark{a}}
}
\startdata
\hline
1 & 10:02:38 & 0.37 \\
2 & 10:13:01 & 0.47 \\
3 & 10:21:56 & 0.56 \\
4 & 10:30:47 & 0.65 \\
5 & 10:39:36 & 0.74 \\
6 & 10:48:28 & 0.83 \\
7 & 10:57:18 & 0.92 \\
8 & 11:06:05 & 0.00 \\
9 & 11:14:53 & 0.09 \\
10 & 11:23:40 & 0.18 \\
\hline
\enddata
\tablenotetext{a}{Determined using the photometric ephemeris given 
in Walker (1965).}
\end{deluxetable}{}

\begin{deluxetable}{ccc}
\tablenum{3}
\tablewidth{4.8in}
\tablecaption{VV Pup Radial Velocity Solution}
\tablehead{
  \colhead{Red-to-Blue phase}
 & \colhead{$\gamma$ velocity (km sec$^{-1}$)}
 & \colhead{K$_2$ (km sec$^{-1}$)}
}
\startdata
\hline
-0.02$\pm0.006$ & -130$\pm18$ & 414$\pm27$ \\
\hline
\enddata
\end{deluxetable}{}

\begin{deluxetable}{cc}
\tablenum{4}
\tablewidth{3.2in}
\tablecaption{Previous High State, Optical K$_2$ Determinations}
\tablehead{
  \colhead{K$_2$ (km sec$^{-1}$)}
 & \colhead{Reference}
}
\startdata
\hline
437\tablenotemark{a} & Schneider \& Young (1980) \\
357$\pm$31 & Cowley et al., (1982) \\
321$\pm$21      & Diaz \& Steiner (1994) \\
467$\pm$40\tablenotemark{b}      & Diaz \& Steiner (1994) \\
\hline
\enddata
\tablenotetext{a}{No uncertainty on this value is given by the authors. They assume
M$_2$ = 0.2 M$_{\odot}$.}
\tablenotetext{b}{Diaz \& Steiner produced this value for K$_2$ by 
applying a K correction to their measured value of 321 km sec$^{-1}$) and
averaging the result with the previous values found by 
Schneider \& Young (1980) and Cowley et al., (1982).}
\end{deluxetable}{}

\begin{deluxetable}{cc}
\tablenum{5}
\tablewidth{2.5in}
\tablecaption{Determined Parameters for VV Pup}
\tablehead{
  \colhead{Parameter}
& \colhead{Value}
}
\startdata
\hline
M$_1$ (M$_{\odot}$) & 0.73$\pm$0.05 \\
M$_2$ (M$_{\odot}$) & 0.10$\pm$0.02 \\
M$_2$ Sp. Type & M6-7V \\
M$_2$ K mag & 15.3$\pm$0.3 \\
M$_2$ M$_K$ & $\sim$10 \\
$i$ & 75$\pm$6$^{\circ}$ \\
Distance & $\sim$120 pc \\
\hline
\enddata
\end{deluxetable}{}

\clearpage

\figcaption[]{Our summed Keck K-band spectrum of VV Pup. 
The eleven single spectra (Table 1) 
were velocity shifted according to our RV solution and co-added to produce this
single spectrum. The y-axis is in relative flux units.
This spectrum is consistent with the secondary star in VV Pup being a M7V star.
}
\figcaption[]{Phase-resolved IRTF/SPEX spectra of VV Pup (see Table 2). 
The spectra are marked with their
orbital phases and shifted
vertically by 0.8, 2.8, 5.0, 6.5, 7.5, 8.5, 9.0, 10.2, and 11.2 $\times 10^{-16}$ in
$\lambda F_{\lambda}$ with respect to the phase 0.00 spectrum. 
Note the appearance of
cyclotron humps in the J-band during phases 0.74-0.09.
}
\figcaption[]{Example Keck/NIRSPEC spectrum. The top panel shows one of our reduced
Keck spectra (orbital phase 0.82) at full resolution. The bottom left box highlights
the Na doublet region while the bottom right shows the same region boxcar 
smoothed by 5.
}
\figcaption[]{Radial velocity measurements for VV Pup (Table 1) using the Na doublet
absorption line. The y-axis gives the measured velocity in km sec$^{-1}$ and the solid curve is our best fit solution listed in Table 3.
}
\figcaption[]{A single IRTF/SPEX spectrum of the LARP EQ Cet. The three obvious 
cyclotron humps at 0.72, 0.96, \& 1.44 microns 
match well to a magnetic field of B=37 MG.
}

\clearpage

\begin{figure}[p!]
\plotone{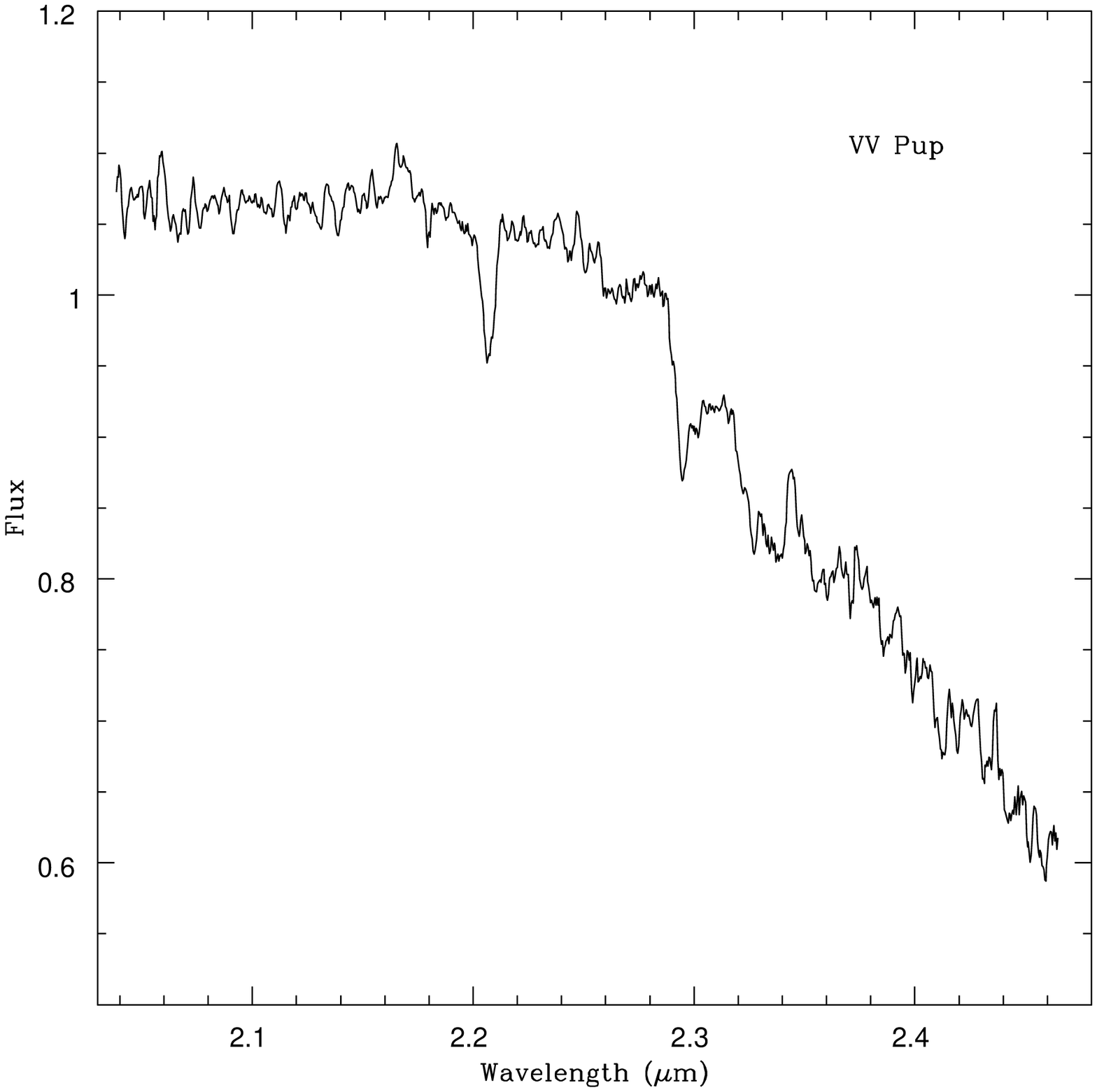}
\end{figure}

\begin{figure}[p!]
\plotone{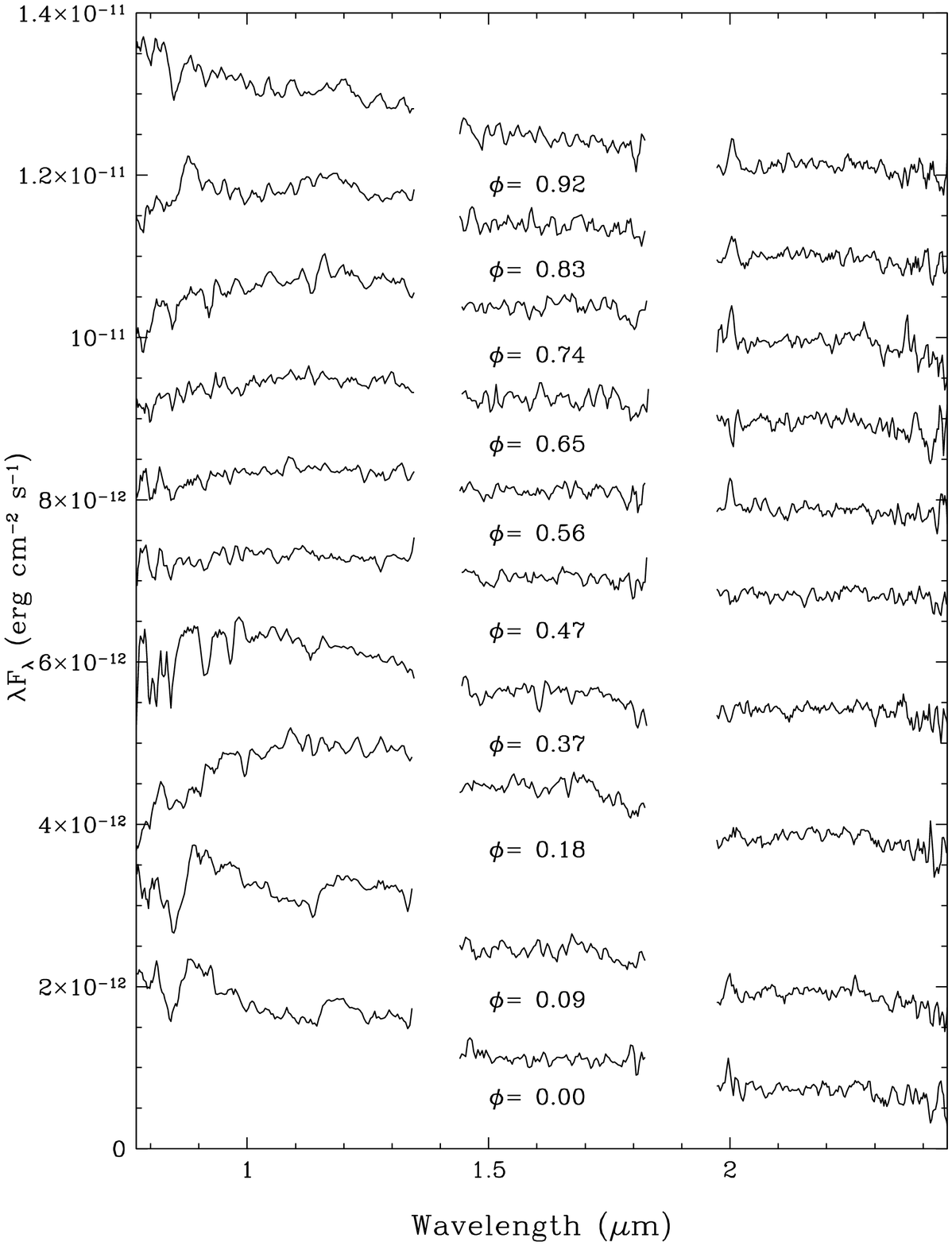}
\end{figure}

\begin{figure}[p!]
\plotone{VVPUP_fig3.ps}
\end{figure}

\begin{figure}[p!]
\plotone{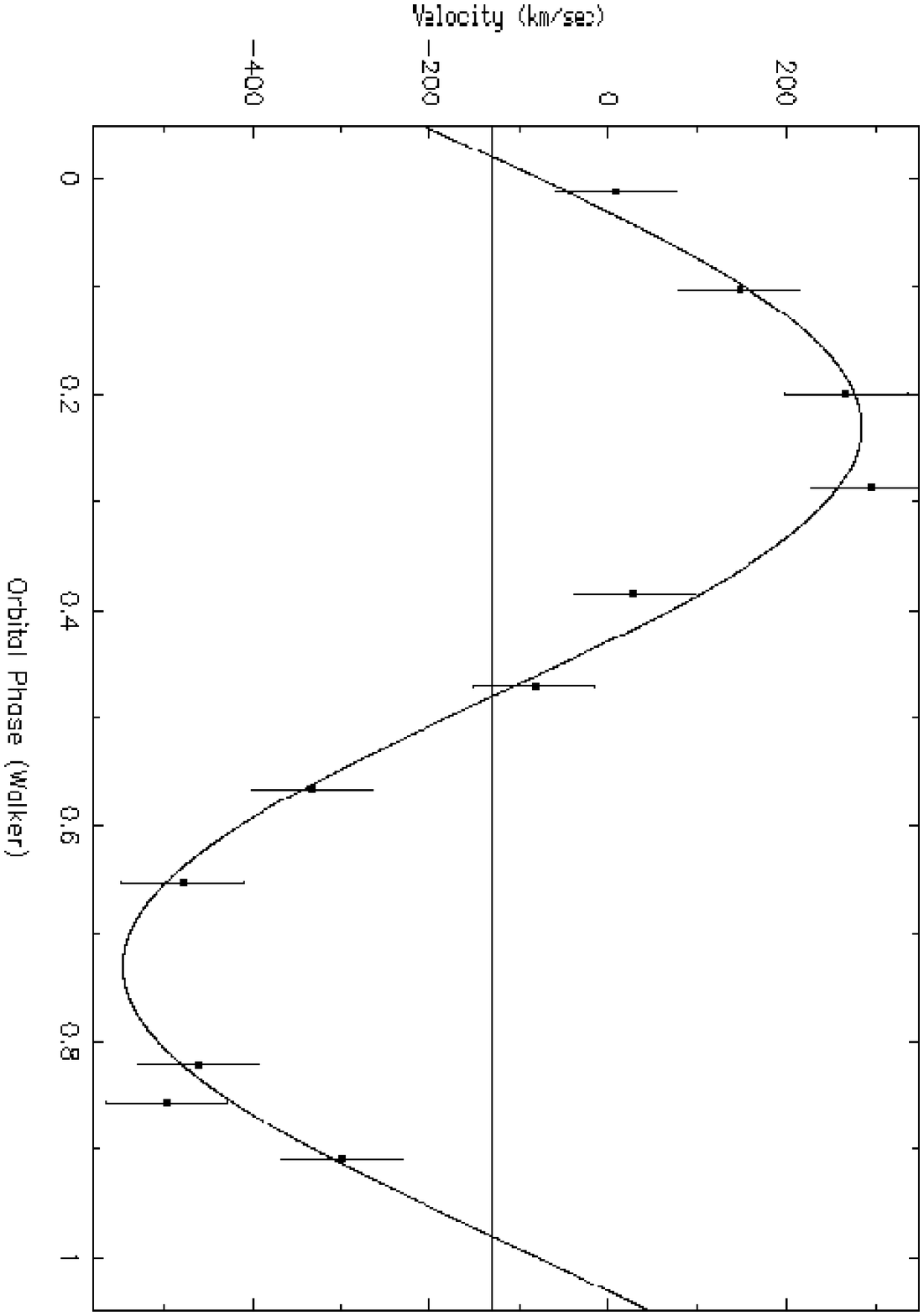}
\end{figure}

\begin{figure}[p!]
\plotone{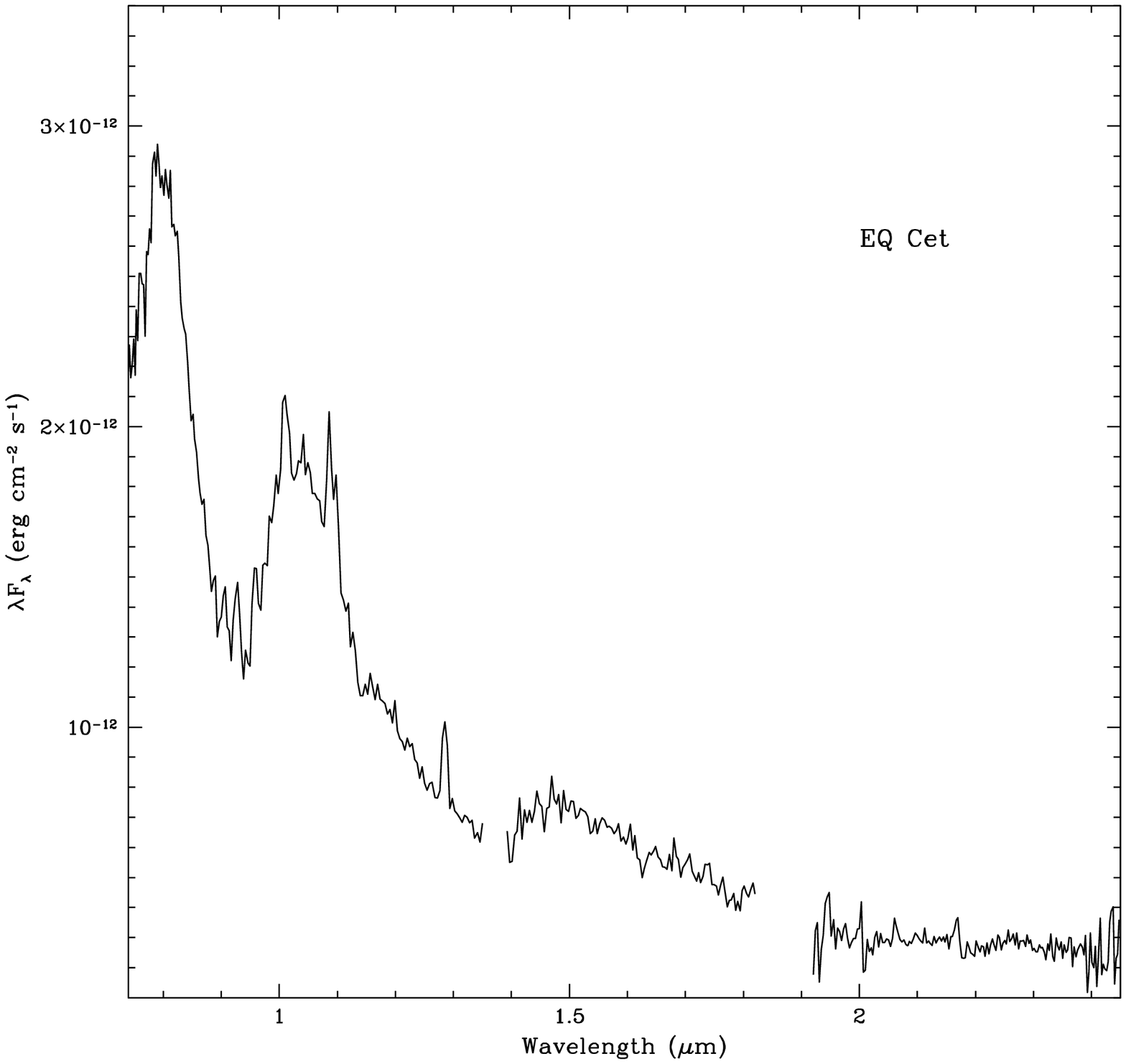}
\end{figure}

\end{document}